\documentclass[10pt]{iopart}

\usepackage{mcite}
\usepackage{amssymb}
\usepackage{bm}
\usepackage{graphicx}
\graphicspath{{./img/}}
\usepackage{hyperref}

\newcommand{\bfsfS}{\mbox{\sffamily{S}}}
\newcommand{\bfsfT}{\mbox{\sffamily{T}}}

\newcommand{\sx}{\bm \sigma_{x}}

\newcommand{\sz}{\bm \sigma_{z}}
\newcommand{\Om}{\Omega}
\newcommand{\be}{\begin{equation}}
\newcommand{\ee}{\end{equation}}
\newcommand{\bea}{\begin{eqnarray}}
\newcommand{\eea}{\end{eqnarray}}

\begin{document}

\title{Landau-Zener transitions in qubits controlled by electromagnetic fields}

\author{Martijn Wubs$^1$,
Keiji Saito$^2$,
Sigmund Kohler$^1$,
Yosuke Kayanuma$^3$ and
Peter H{\"a}nggi$^1$}

\address{$^1$ Institut f{\"u}r Physik, Universit{\"a}t Augsburg,
Universit{\"a}tsstra{\ss}e 1, D-86135 Augsburg, Germany}

\address{$^2$ Department of Physics, Graduate School of Science,
University of Tokyo, Bunkyo-Ku, Tokyo 113-0033, Japan}

\address{$^3$ Department of Mathematical Science, Graduate School of
Engineering, Osaka Prefecture University, Sakai 599-8531, Japan}

\ead{Martijn.Wubs@Physik.Uni-Augsburg.DE}

\begin{abstract}
We investigate the influence of a dipole interaction with a
classical radiation field on a qubit during a continuous change of a
control parameter. In particular, we explore the non-adiabatic
transitions that occur when the qubit is swept with linear speed
through resonances with the time-dependent interaction. Two
classical problems come together in this model: the Landau-Zener
 and the Rabi problem. The probability of Landau-Zener
transitions now depends sensitively on the amplitude, the frequency
and the phase of the Rabi interaction. The influence of the static
phase turns out to be particularly strong, since this parameter
controls the time-reversal symmetry of the Hamiltonian. In the
limits of large and small frequencies, analytical results obtained
within a rotating-wave approximation compare favourably with a
numerically exact solution. Some physical realizations of the model
are discussed, both in microwave optics and in magnetic systems.

\end{abstract}




\section{Introduction}

An essential ingredient to a quantum computer is a set of parameters
that is controllable in the sense that it is possible to manipulate
the parameter values at any time such that the qubits undergo
one-qubit or two-qubit gate operations.  For quantum computer
implementations that rely on nuclear magnetic resonance
\cite{Nielsen00} or on spins in quantum dots \cite{Loss1998a}, such
a manipulation is possible by switching magnetic fields that act on
the qubit.  This includes the possibility of inverting the sign of
an acting magnetic field. As the field changes sign, the diabatic
energy levels of the qubit typically cross. If at the same time a
second magnetic field acts in any other direction, the adiabatic
levels form an \textit{avoided} crossing instead of an exact
crossing. Then, depending on the speed at which the control
parameters are manipulated, the state of the qubit can follow the
adiabatic energy levels or undergoes a non-adiabatic, so-called
Landau-Zener (LZ) transition to the opposite branch
\cite{Landau32,Zener32,Stueckelberg32,Kayanuma84,Grifoni98}.

In the context of quantum computation, it has been proposed to
exploit LZ transitions for improving the readout of qubits via the
so-called Zener flip quantum tunneling \cite{Ankerhold03}. This
mechanism has recently been implemented for flux qubits
\cite{Ithier05}. A method for non-adiabatic electron manipulation in
quantum dots also relies on LZ transitions \cite{Saito04}. Moreover,
the observation of LZ transitions is a clear sign of coherence like,
e.g., optical coherence in a classical optical ring resonator
\cite{Spreeuw90} or macroscopic quantum coherence in superconducting
loops \cite{Shytov03,Izmalkov04}. LZ transitions have also been used
to determine tiny interactions between levels in molecular clusters
\cite{Wernsdorfer99}. While in these cases, LZ transitions are
beneficial, the opposite is true in the case of adiabatic quantum
computing \cite{Farhi00, Steffen03}. There, the computation is
performed by a quantum system that follows adiabatically the
instantaneous ground state of a slowly varying Hamiltonian and,
consequently, the emergence of any non-adiabatic transition
constitutes an error source.

The physical origin of a coupling between two levels of a quantum
system is not necessarily simply an overlap between the respective
wave functions. In particular for spins and atoms, such a coupling
typically stems from the dipole interaction of the system with a
radiation field. In a seminal work \cite{Rabi37a} (see also
\cite{Schwinger37}), Rabi predicted within an exact quantum
mechanical treatment that a classical, monochromatic and circularly
polarized radiation field induces spin rotations with a frequency
which at resonance is proportional to the field amplitude. With
resonant linearly polarized light, the same characteristic harmonic
Rabi oscillations of atomic inversions are observed.  As a
consequence of the linearly polarized driving, the optical
realization of the Rabi problem is not exactly solvable. For
resonant excitations, however, it is possible to apply a
rotating-wave approximation (RWA) which formally restores the
situation with circular polarization \cite{Allen75,Mandel95}.  This
necessarily neglects effects beyond RWA like the Bloch-Siegert shift
of the resonance frequency.  In optical realizations, however, such
these effects are very tiny \cite{Bloch1940a,Allen75}.

The question now arises whether a level interaction mediated by a
classical monochromatic  radiation field can induce Landau-Zener
transitions in a two-level system as its energies cross. In this
work, we demonstrate that this is indeed the case.  Thereby, we
investigate LZ transitions that are induced by the coupling of a
spin to a linearly polarized light field, henceforth referred to as
Rabi coupling.  In the traditional LZ problem, non-adiabatic
transitions occur when the adiabatic energy levels are close to each
other.  By contrast, we will find that with a Rabi coupling to a
high-frequency field, the transitions take place at times at which
the radiation field is at resonance with the diabatic energy levels.
This allows for sufficiently weak coupling a perturbative treatment
within a rotating-wave approximation. For suitably chosen
parameters, the driving reduces the probability for LZ transitions
which relates this problem to the so-called coherent destruction of
tunneling \cite{Grossmann91a,Grossmann91b}.
A different kind of time-dependence would be provided by coupling
the two-level system to a noise source. In this paper, we will work
in the coherent limit instead, which at least in possible
realizations of quantum computation should be a good approximation
on a relevant time scale. In relation to this, it is interesting to
note that LZ tunneling is fairly robust against classical noise
\cite{Kayanuma84,Kayanuma85,Nishino01,Pokrovsky03} and quantum
dissipation
\cite{Gefen87,Ao91,Shimsony91,Akulin92,Kayanuma98,Shytov00,Saito02}.

In this context, we like to emphasize that herein considered
Landau-Zener tunneling due to coupling to a light field is different
from the one considered in Refs.~\cite{Kayanuma94,
Garraway97,Izmalkov04} where it is the diabatic energies of the
two-level system that are subject to a time-periodic modulation (the
so-called dynamic Stark effect).  Still, the model considered here
applies quite generally to qubits whose level interaction varies
harmonically in time due to interaction with an external field that
causes a negligible dynamic Stark effect. In the discussion at the
end of the paper some possible physical realizations of the model
are suggested.
%

\section{The Landau-Zener model with harmonic interaction modulation}
\label{sec:model}

We consider a quantum system (``atom'') with two relevant energy
levels $|1\rangle$ and $|2\rangle$ whose time-dependent energies
$\pm Vt/2$ cross at $t=0$.  Both levels are coupled to a classical
dipole field with frequency $\Omega$ and phase $\phi$.  The effective
amplitude $g$ is given by field strength times the dipole moment of
the two-level system.  Thus, the Hamiltonian reads
\begin{equation}
\label{Ht} H(t) = \frac{V t}{2}\sz + f(t) \sx , \quad f(t) =
g\cos(\Omega t+\phi),
\end{equation}
where $\sz|1\rangle = |1\rangle$ and $\sz|2\rangle = -|2\rangle$.
Moreover, we assume that at time $t=-\infty$, the system is in its
instantaneous ground state, i.e.\ $|\psi(-\infty)\rangle =
|1\rangle$.  If the energies in the first term would be
time-independent, and if the field would be in resonance with the
atomic energy difference, then this Hamiltonian would describe an
atom undergoing Rabi oscillations.

Since for most times, the Hamiltonian (\ref{Ht}) is dominated by its
first term, a proper interaction-picture representation is provided
by the transformation $U_{0}(t) = \exp(- i Vt^2\sz/2\hbar)$, that is
$|\psi(t)\rangle = U_{0}(t)|\tilde{\psi}(t)\rangle$ and
$|\tilde{\psi}(t)\rangle=[ \tilde{c}_1(t)|1\rangle +
\tilde{c}_2(t)|2\rangle]$, where the interaction-picture probability
amplitudes obey
\begin{equation}
\label{eqmotint}
\left(\begin{array}{c}\dot{\tilde{c}}_{1} \\ \dot{\tilde{c}}_{2} \end{array} \right)
=  - \frac{\mathrm{i}}{\hbar}
  \left(\begin{array}{cc} 0 & f(t)\;\e^{\mathrm{i} V t^{2}/2\hbar}  \\
  f(t)\;\e^{-\mathrm{i} V t^{2}/2\hbar} & 0 \end{array}\right)
  \left(\begin{array}{c}\tilde{c}_{1} \\ \tilde{c}_{2} \end{array} \right) .
\end{equation}

For $\Omega=\phi=0$, the Hamiltonian (\ref{Ht}) defines the standard
Landau-Zener problem for which the exact solution of the equation of
motion (\ref{eqmotint}) can be expressed in terms of parabolic
cylinder functions \cite{Landau32}.  Then, the time-evolution from
$t=-\infty$ to $t=\infty$ is given by the S-matrix
\begin{equation}\label{transitionS}
\bfsfS_g = \left( \begin{array}{cc} \sqrt{q} & \sqrt{1-q}\;\e^{-\mathrm{i}\chi}
\\- \sqrt{1-q}\;\e^{\mathrm{i} \chi} & \sqrt{q} \end{array}\right) ,
\end{equation}
where $q = \exp(-2\pi g^2/\hbar V)$ and the Stokes phase $\chi =
\pi/4+\arg\Gamma(1-\mathrm{i}\delta)+\delta(\ln \delta-1)$, with
$\delta = g^{2}/(\hbar V)$ and $\Gamma(..)$ the Gamma function.
The famous Landau-Zener transition probability follows readily: the
probability $P$ that the atom ends up in the initially unoccupied level
$|2\rangle$ is
\begin{equation}
 \label{standardpLZ}
 P\equiv  |c_{2}(t=\infty)|^2 = 1 - \e^{-2\pi g^{2}/\hbar V}.
\end{equation}
Note that this result is exact for all values of $g$ and $V$.

Already a good approximation to the time-dependent solution is
provided by the fact that for $f(t)=g$ and sufficiently large times,
the phase factors in the matrix in Eq.~(\ref{eqmotint}) are rapidly
oscillating with a quadratic time dependence.  As a consequence,
$\tilde{c}_1$ and $\tilde{c}_2$ remain essentially constant.  By
contrast, close to $t=0$ these phase factors assume a stationary
value and the two-level system undergoes a transition.  This means
that S-matrix (\ref{transitionS}) in fact describes a transition
taking place at $t=0$.  Thus, in the interaction picture the
dynamics is approximately given by $|\tilde{\psi}(t)\rangle =
|\tilde{\psi}(-\infty)\rangle$ for $t<0$ and
$|\tilde{\psi}(t)\rangle = \bfsfS_g|\tilde{\psi}(\-\infty)\rangle$
for $t>0$.

\begin{figure*}[t]
\centerline{\includegraphics{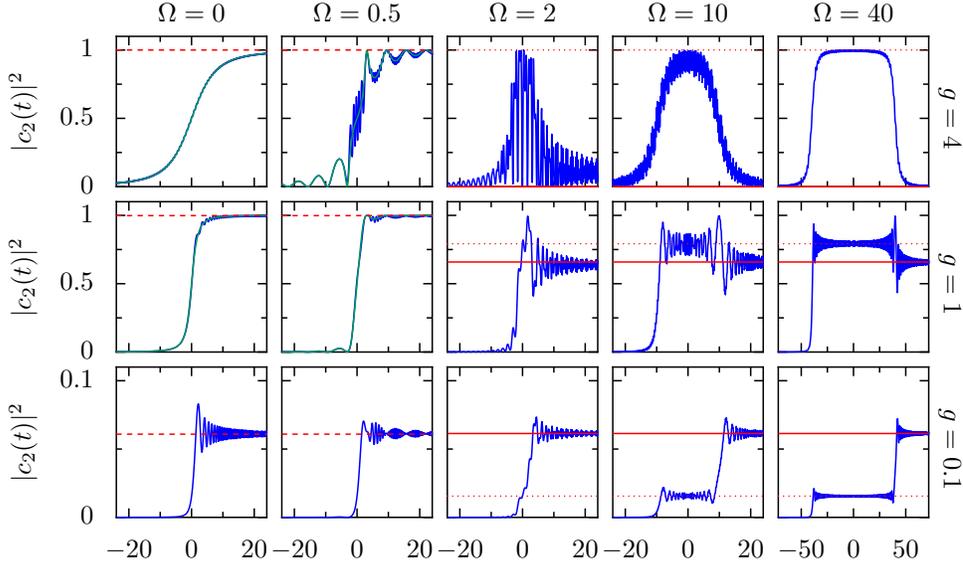}} \caption{LZ
transition probability $|c_{2}(t)|^2$ as a function of time in units
of $(\hbar/V)^{1/2}$ for $\phi=0$ and various values of the
interaction strength $g$ in units $(\hbar V)^{1/2}$ and the
modulation frequency $\Omega$ in units of $(V/\hbar)^{1/2}$.  The
red lines mark the standard LZ transition probability
[Eq.~(\ref{standardpLZ}), dashed], the RWA result for a double
transition [Eq.~(\ref{PLZ0}), solid], and the transition probability
at the intermediate stage [Eq.~(\ref{Pint}), dotted]. The green
curves in the four upper-left panels  correspond to the
adiabatic-following result Eq.~(\ref{p2adiabatic}).}
\label{timeplots}
\end{figure*}

For $\Omega\neq 0$, the time $t=0$ no longer marks the time at which
the phase is stationary and the behaviour changes significantly, as
we will see below.  In order to anticipate the richness of the
resulting dynamics, we have numerically integrated the equations of
motion~(\ref{eqmotint}) for $\phi=0$ and various coupling strengths
$g$ and frequencies $\Omega$. Fig.~\ref{timeplots} depicts the time
dependence of the probability $|c_{2}(t)|^{2}$ to find the atom in
the initially unpopulated level $|2\rangle$.
For small interaction $g$ (see lower five plots in
Fig.~\ref{timeplots}), the final transition probability for long
times does not depend strongly on frequency, although  the curves
differ strongly around $t=0$. The most interesting feature of
Fig.~\ref{timeplots} is that for high frequencies, $\Omega \gg
\sqrt{V/\hbar}$, the dynamics consists of two (almost) independent
transitions.

\section{Adiabatic vs.\ non-adiabatic transitions}

In the standard Landau-Zener problem, one distinguishes two limiting
cases: If the level crossing occurs very rapidly, the potential
switches practically instantaneously such that no significant
dynamics can take place.  The system will then  remain in level
$|1\rangle$ so that finally $P=0$. In the opposite limit the
instantaneous energy levels change very slowly. The system then
follows adiabatically the lower-energy level $|E_{-}(t)\rangle$ and
ends up with $P=1$.

\subsection{Adiabatic following}
Adiabatic following means that transitions between the instantaneous
eigenstates can be neglected. The criterion for adiabatic following
is that at each instance of time, the coupling between the adiabatic
energies is ``sufficiently small'', much smaller than the energy
splitting. Stated in mathematical terms, this requirement becomes
$|\langle\; E_{-}(t)|\frac{\mbox{d}}{\mbox{d}t}|E_{+}(t)\;\rangle |
\ll |E_{+}(t)-E_{-}(t)|/\hbar$. This gives for the standard
Landau-Zener problem  a splitting $2g$ and the adiabaticity
condition $\hbar V \ll g^2$.

\begin{figure}[t]
\centerline{\includegraphics{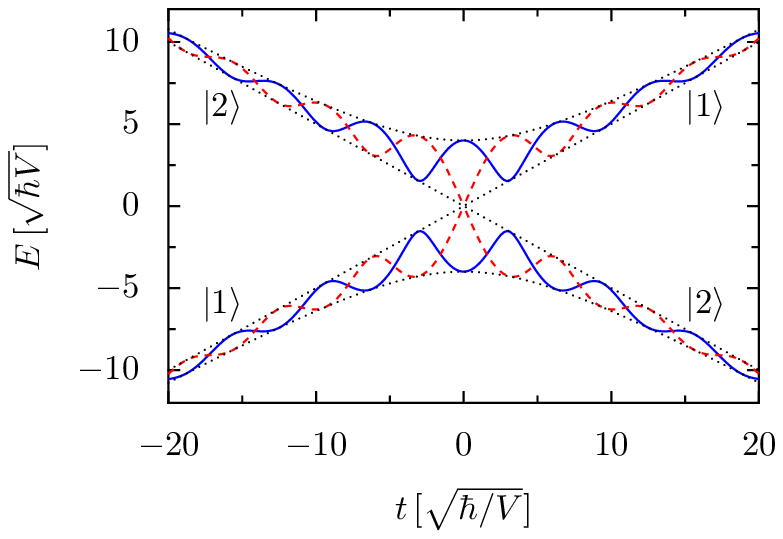}}
\caption{Adiabatic energies $E_{\pm}(t)$ for a Rabi
coupling with $\phi=0$ (solid lines) and $\phi=\pi/2$ (dashed).
The dotted lines marked the limiting values for $g=0$ and the standard
LZ model with constant coupling, respectively.  The other parameters
are $g=4\sqrt{\hbar V}$, and $\Omega = 0.5\sqrt{V/\hbar}$.}
\label{adiabaticenergies}
\end{figure}

For the time-dependent two-state Hamiltonian (\ref{Ht}), the
condition for adiabatic following becomes more involved, because the
minimal splitting of the adiabatic energies $
E_{\pm}(t)=\pm\sqrt{(Vt/2)^2 + f^{2}(t)}$ depends not only on the
coupling strength $g$ but also on the frequency $\Omega$ and the
phase $\phi$. The minimal splitting will never be larger than $2g$,
whatever the frequency and phase. The sensitive dependence on the
phase becomes obvious from Fig.~\ref{adiabaticenergies}: In
particular for $\phi=\pi/2$, the adiabatic spectrum no longer
exhibits an avoided crossing but rather an exact crossing, since at
time $t=0$ both terms in the Hamiltonian (\ref{Ht}) vanish
simultaneously. Consequently, the adiabaticity condition is violated
irrespective of the values of the $g$ and $\Omega$.  This
qualitative difference already provides a hint that the phase $\phi$
has a strong influence on the population dynamics.

By contrast, for a phase $\phi\neq \pi/2
(\mathop{\mathrm{mod}}\pi)$, the energies $E_\pm$ never form an
exact crossing. Thus, it is always possible to choose $V$ and
$\Omega$ so small that the adiabaticity condition is fulfilled.
Then, by making the time-dependent transformation to the
instantaneous-energy representation,  neglecting there the
off-diagonal elements in the equations of motion, solving the
dynamics and then transforming back to the diabatic representation,
it follows that the probability $|c_{2}(t)|^{2}$ to find the qubit
in diabatic state $|2\rangle$ goes from zero to one as
\begin{equation}
\label{p2adiabatic} |c_{2,{\mathrm adiabatic}}(t)|^{2} =
\frac{1}{2}\Bigg( 1
  + \frac{V t}{\sqrt{\left({V t}\right)^{2} + 4f^{2}(t)}}\Bigg).
\end{equation}
The dependence for intermediate times on the interaction modulation
$f(t)$ is clearly seen.  The four upper-left
 panels in Fig.~\ref{timeplots} with $\Omega = 0$ or $0.5$ and $g=1$ or $4$ are very well
described by the adiabatic-following result~(\ref{p2adiabatic}).

\subsection{Non-adiabatic regime}
\label{sec:nonadiabatic}

For phase $\phi=0$, the Rabi coupling $f(t)$ is zero at time
$t=\pi/2\Omega$ and the energy splitting becomes $\pi V/2\Omega$.
This means that for a large driving frequency $\Omega >
\sqrt{V/\hbar}$, the adiabaticity condition is violated. The data
shown in the right columns of Fig.~\ref{timeplots} indicate that in
this regime the dynamics consists of two transitions at times
$\mp\hbar\Omega/V$.  If the time $2\hbar\Omega/V$ between the
individual transitions is sufficiently large, as specified below,
the two transitions are essentially independent of each other. Then,
it is possible to derive within a rotating-wave approximation an
analytical expression for the final transition probability.  The
derivation is closely related to the transfer matrix method employed
in Refs.~\cite{Kayanuma93,Kayanuma94}.

With the new variables $d_{1}(t)={\tilde c}_{1}\exp(\mathrm{i}
\hbar\Omega^{2}/4V)$ and $d_{2}(t)={\tilde
c}_{2}\exp(-\mathrm{i}\hbar \Omega^{2}/4V)$, one obtains from
Eq.~(\ref{eqmotint}) the equations of motion
\begin{eqnarray}
\dot{d}_{1}
&=& -\mathrm{i}\frac{g}{2\hbar}
    \left[ \e^{\mathrm{i} V (t+\hbar\Omega/V)^{2}/2\hbar}
          +\e^{\mathrm{i} V (t-\hbar\Omega/V)^{2}/2\hbar}\right]d_{2},
   \label{eqmotintford12_a} \\
\dot{d}_{2}
&=& -\mathrm{i}\frac{g}{2\hbar}
    \left[ \e^{-\mathrm{i} V (t+\hbar\Omega/V)^{2}/2\hbar}
          +\e^{-\mathrm{i} V (t-\hbar\Omega/V)^{2}/2\hbar}\right]d_{1}.
   \label{eqmotintford12_b}
\end{eqnarray}
Like in Eq.~(\ref{eqmotint}), the phases on the right-hand side obey
a quadratic time dependence.  Thus, with the arguments provided
after Eq.~(\ref{transitionS}), we can conclude that each phase
factor is relevant only at times at which the phase is stationary,
i.e., the first term contributes only at time $t_-=-\hbar\Omega/V$
while the second term becomes relevant at time $t_+=\hbar\Omega/V$.
Thus, we keep at both times $t_-$ and $t_+$ only the respective
resonant term while the ``counter-rotating'' term is neglected. (To
be sure, two separate rotating-wave approximations are needed
corresponding to $t_{+}$ and $t_{-}$.) Then, at times close to
$t_\mp$, the equation of motion is of the same form as
Eq.~(\ref{eqmotint}) and the dynamics is determined by the
S-matrix~(\ref{transitionS}) with $g$ replaced by $g/2$, i.e.\
$\bfsfS_\mp = \bfsfS_{g/2}$. Consequently within the transfer matrix
approximation, the time evolution becomes
\begin{equation}
|\tilde{\psi}(t)\rangle = \left\{\begin{array}{lcl}
    |\psi(-\infty)\rangle  &\mbox{for} & t<-\hbar\Omega/V,
  \\[1ex]
    \bfsfS_{g/2}|\psi(-\infty)\rangle  &\mbox{for} &
    -\hbar\Omega/V<t<\hbar\Omega/V,
  \\[1ex]
    \bfsfS_{g/2}^2|\psi(-\infty)\rangle  &\mbox{for} &
    t>\hbar\Omega/V.
  \end{array}\right.
\end{equation}
With this expression, the probability to find the system at time
$t=\infty$ in state $|2\rangle$ is readily evaluated as
\begin{equation}
\label{PLZ0}
P = |\langle 2|\bfsfS_{g/2}^2|1\rangle|^2
  = 4\; \e^{-\pi g^2/2\hbar V} \big(1- \e^{-\pi g^2/2\hbar V}\big) .
\end{equation}
During the intermediate times $-\hbar\Omega/V< t <\hbar\Omega/V$,
the occupation probability of level $|2\rangle$ becomes \be
P_\mathrm{int} = |\langle 2|\bfsfS_{g/2}|1\rangle|^2 = 1-\exp(-\pi
g^2/2\hbar V). \label{Pint} \ee Note that in these expressions, the
exponent differs from the exponent in Eq.~(\ref{standardpLZ}) by a
factor $1/4$. Moreover, $P$ in Eq.~(\ref{PLZ0}) no longer depends
monotonously on the coupling strength as in the standard LZ problem,
but rather assumes a maximum for $\exp(-\pi g^2/2\hbar V) =
\frac{1}{2}$.
%
Interestingly enough, the transfer matrix results are independent of
the Stokes phase $\chi$ and the modulation frequency $\Omega$.  The
independence of the frequency  is confirmed by Fig.~\ref{timeplots},
where  the transfer-matrix results (red lines) nicely agree with the
exact results (blue lines) for $\Omega=2$, $10$, and $40$. Clearly,
at long times, the probability $|c_2|^2$ is a function of only the
coupling strength $g$.

%

\begin{figure}[t]
\centerline{\includegraphics{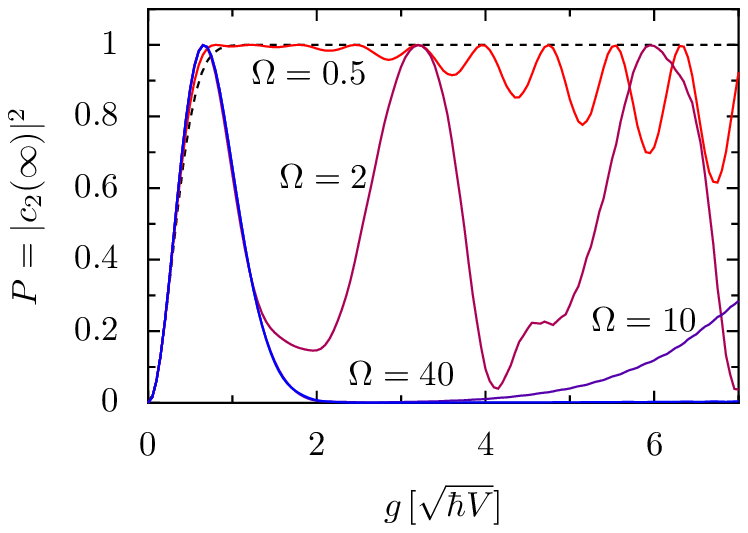}} \caption{Final
transition probability $P$ as a function of coupling $g$, for
several values of the interaction modulation frequency $\Omega$. The
dashed line marks the standard LZ transition
probability~(\ref{standardpLZ}) valid for $\Omega=0$. The transfer
matrix result (\ref{PLZ0}) coincides with the numerical result for
$\Omega=40$. (The numerical time integration was performed from
times $-500 \sqrt{\hbar/V}$ to $+500 \sqrt{\hbar/V}$.)}
\label{gplots}
\end{figure}

Below we will compare more systematically the transition
probabilities obtained from the transfer matrix method with a
numerically exact solution. But first we have to specify the
conditions under which the time between two LZ transitions will be
long enough for the transfer-matrix analysis to hold. It should be
remembered that LZ transitions neither occur instantaneously nor
take infinitely long \cite{Mullen89,Vitanov96,Vitanov99}. The time
$2\hbar\Omega/V$ between the two consecutive transitions should be
larger than the duration of a single transition. It has been
estimated that the standard LZ transition has a typical duration
$\tau_{\mathrm{LZ}} \simeq \sqrt{\hbar/V}$ when non-adiabatic
transitions are probable ($2 g^2/(\hbar V) \ll 1$), while
$\tau_{\mathrm LZ}\simeq 2 g/V$ in the adiabatic limit $2 g^2/(\hbar
V) \gg 1$ \cite{Mullen89}. Correspondingly, reliable results of the
transfer matrix approach are to be expected if \bea \Omega & \gtrsim
&  \sqrt{V/(4\hbar)} \qquad
\mbox{for} \quad g^2/(2\hbar V) \ll 1,  \label{firstcondition} \\
\Omega &\gtrsim & g/\hbar \qquad\qquad \mbox{for} \quad g^2/(2\hbar
V)\gg 1. \label{secondcondition}\eea These estimates are confirmed
by the numerical solution of Eq.~(\ref{eqmotint}) as plotted in
Fig.~\ref{gplots}, which depicts the probability of finding the
system in state $|2\rangle$ at large times. The figure makes clear
that for small coupling strengths $g\ll \sqrt{\hbar V}$ the
condition~(\ref{firstcondition}) is sufficient but not necessary,
because Eq.~(\ref{PLZ0}) is seen to be accurate irrespective of the
frequency. This is in accordance with the  lower five panels in
Fig.~\ref{timeplots} and with fact that the standard LZ
result~(\ref{standardpLZ}) and the transfer-matrix
expression~(\ref{PLZ0})  agree  that $P$ equals $2\pi g^2/V$ to
first order in $g^2/V$. On the other hand, we  find significant
deviations from the expression~(\ref{PLZ0}) once $g$ becomes larger
than $\sqrt{\hbar V}$ and of the order $\hbar\Omega$. This is where
the two LZ transitions start to ``feel'' each other.  For
sufficiently large coupling, the transition probability even
increases again and assumes further maxima with $P=1$. It is nice
that the argument can be turned around and that LZ times can be
estimated with the help of the frequencies at which the exact and
the transfer-matrix results start to deviate. In doing so, we indeed
find (here and in Sec.~\ref{crossover}) that $\tau_{\mathrm
LZ}\simeq 2 g/V$ for $2 g^2/(\hbar V) \gg 1$, in agreement with
\cite{Mullen89}. The present model provides an independent and
simple method to determine Landau-Zener times.

\section{Exploring the crossover region}\label{crossover}

Our analysis has identified two different parameter regimes in which
the analytical solution is confirmed by the our numerical results.
First, there is the regime of slow driving in which $\hbar\Omega$
denotes the smallest energy scale of the problem.  Then, the time
dependence of the coupling is not essential and the transition
probabilities are the same as in the standard LZ problem. In the
second regime $\hbar\Omega$ is the largest energy scale and the
transfer matrix results hold. In particular, we find $P=1$ for $\pi
g^2=2\hbar V\ln 2$.
In this section, we complement our analytical findings by numerical
results for the intermediate parameter regime.

\begin{figure}[t]
\centerline{\includegraphics{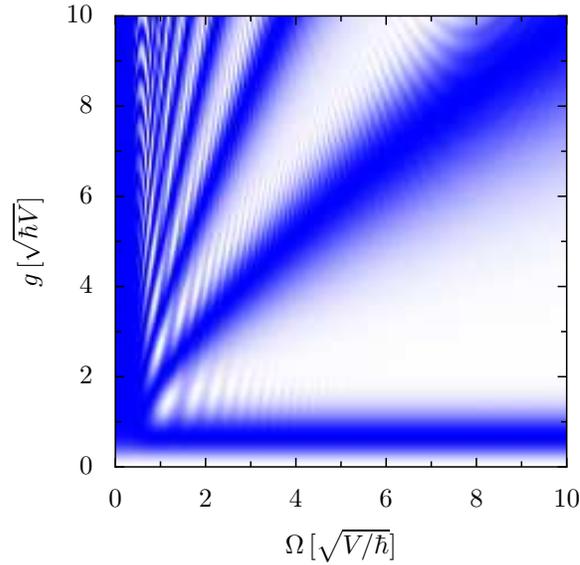}}
\caption{Final transition probability $P$ as a function of coupling
strength $g$ and frequency $\Omega$.
Blue areas correspond to $P=1$, white areas to $P=0$.}
\label{gWplots}
\end{figure}

Figure~\ref{gWplots} shows the final transition probability, that is
the occupation of state $|2\rangle$ in the limit $t\to\infty$. (The
time interval for numerical integration was chosen the same as for
Fig.~\ref{gplots}.)  The vertical stripe with $P=1$ for
$\Omega\lesssim 0.5$ corresponds to the adiabatic regime. The
horizontal blue stripe marks the maximum found within the transfer
matrix approach. The figure also confirms (i) that the location of
the maximum has no significant frequency dependence and (ii) that
$P$ decays for a larger coupling $g$ almost to zero, yielding the
white region with $P=0$ above the horizontal blue band, in agreement
with the transfer-matrix prediction~(\ref{PLZ0}). Increasing $g$
further, we find that at $g\approx\hbar\Omega$, the transition
probability again assumes values close to unity.  This regime,
including the sequence of maxima and minima that can be observed for
even larger coupling, is beyond the range of validity of the
transfer matrix method. The fact that the transfer matrix approach
starts to break down along the diagonal $g=\hbar \Omega$ in
Figure~\ref{gWplots} neatly agrees with the
estimate~(\ref{secondcondition}). With the reasoning given in
Sec.~\ref{sec:nonadiabatic}, we can infer from Fig.~\ref{gWplots}
that the estimate for the LZ time $\tau_{\mathrm LZ}\simeq 2 g/V$
\cite{Mullen89} holds at least in the broad parameter regime $1 <
g^{2}/\hbar V < 100$.

\section{Phase dependence}

When discussing the adiabatic energies of the Hamiltonian
(\ref{Ht}), we have already anticipated that the phase $\phi$ might
have some relevance that we explore in the following.  For that
purpose, we adapt the analytical approach of
Sec.~\ref{sec:nonadiabatic} accordingly.

\subsection{Transfer matrix approach}

Inserting again the  definitions $d_1 = c_1
\exp(\mathrm{i}\hbar\Omega^2/4V)$ and $d_2 = c_2
\exp(\mathrm{i}\hbar\Omega^2/4V)$ into the equation of motion
(\ref{eqmotint}), we find
\begin{eqnarray}
\dot{d}_{1}
&=& -\mathrm{i}\frac{g}{2\hbar}
    \left[ \e^{\mathrm{i} V (t+\hbar\Omega/V)^{2}/2\hbar+\mathrm{i}\phi}
          +\e^{\mathrm{i} V (t-\hbar\Omega/V)^{2}/2\hbar-\mathrm{i}\phi}
   \right] d_{2},
   \label{eqmotintapproxtransphi_a} \\
\dot{d}_{2}
&=& -\mathrm{i}\frac{g}{2\hbar}
    \left[ \e^{-\mathrm{i} V (t+\hbar\Omega/V)^{2}/2\hbar-\mathrm{i}\phi}
          +\e^{-\mathrm{i} V (t-\hbar\Omega/V)^{2}/2\hbar+\mathrm{i}\phi}
   \right] d_{1}.
   \label{eqmotintapproxtransphi_b}
\end{eqnarray}
These equations differ from Eq.~(\ref{eqmotint}) merely by the phase
$\phi$ in the exponents.  The goal is now to transform
Eqs.~(\ref{eqmotintapproxtransphi_a}) and
(\ref{eqmotintapproxtransphi_b}) such that they assume at times
$t_\mp = \mp \hbar\Omega/V$ the same form as
Eqs.~(\ref{eqmotintford12_a}) and (\ref{eqmotintford12_b}). After
such a transformation the transfer matrix method could be used
again. At time $t_-$, when only the first term in the equations of
motion is relevant, an appropriate transformation reads
\begin{equation}
\bfsfT =
  \left(\begin{array}{cc} \e^{-\mathrm{i}\phi/2} & 0  \\
  0 & \e^{\mathrm{i}\phi/2} \end{array}\right) .
\end{equation}
The corresponding transfer matrix $\bfsfS_-$ follows from a
transformation of $\bfsfS_{g/2}$ with $\bfsfT$ and reads $\bfsfS_- =
\bfsfT^{-1} \bfsfS_{g/2} \bfsfT$.
With the same reasoning, we find that at time $t_+$ the required
transformation is $\bfsfT^{-1}$ and the transfer matrix is $\bfsfS_+
= \bfsfT \bfsfS_{g/2} \bfsfT^{-1}$.  Consequently, the complete time
evolution becomes $|\tilde{\psi}(\infty)\rangle =  \bfsfS_+ \bfsfS_-
|\tilde{\psi}(-\infty)\rangle$. This leads to the final transition
probability
\begin{equation}
\label{p2withphi}
 P = 4\;\e^{-\pi g^{2}/2V} \left(1-\e^{-2\pi g^{2}/2V}\right)\cos^{2}\phi.
\end{equation}
Clearly,  the phase shift modifies the transition probability by a
factor $\cos^2\phi$.  At intermediate times $t\approx 0$, the
occupation is determined by  $ \bfsfS_-$. Interestingly enough, the
absolute values of its matrix elements do not depend on $\phi$ and,
thus, we find at $t=0$ the same phase-independent transition
probability $P_{\mathrm int}$ as in (\ref{Pint}). Evidently, the
phase dependence in~(\ref{p2withphi}) is caused by quantum
interference between the two transition paths from $|1\rangle$ to
$|2\rangle$, which also explains the absence of any phase dependence
after the first transition.

\begin{figure}[t]
\begin{center}
\centerline{\includegraphics{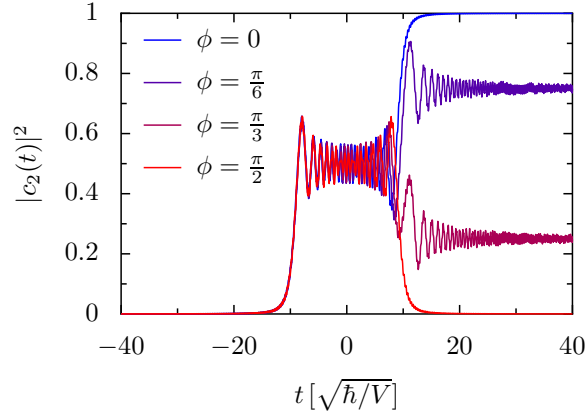}}
\end{center}
\caption{Transition probability as a function of time for
$g/\sqrt{\hbar V}= [2\ln{(2)/\pi]^{1/2}} \approx 0.664$ and driving
frequency $\Omega=10\sqrt{V/\hbar}$.  The transfer matrix method
predicts the intermediate population $|c_2(0)|^2 = 0.5$ and the
final transition probabilities 1, 0.75, 0.25, and 0, respectively.}
\label{phiplot}
\end{figure}

Figure~\ref{phiplot} shows a comparison with the
numerically computed time evolution.  The long-time limits compare
favourably with our prediction for the final state.
As for the special case $\phi=0$, the final occupation is independent
of the frequency, provided that $\Omega$ exceeds $g/\hbar$ and
$\sqrt{V/\hbar}$.
The strong phase dependence in~(\ref{p2withphi}) may come as a
surprise, since it is tempting to argue that for high frequencies,
phase relations should be immaterial due to the many oscillations
occurring during each LZ transition.  However, figure~\ref{phiplot}
clearly demonstrates that such  reasoning is incorrect.

\subsection{Time-reversal anti-symmetry}

The transition probability (\ref{p2withphi}) obviously vanishes for
$\phi=\pi/2$.  This behaviour can already been obtained from
symmetry arguments.  For this phase, the Hamiltonian (\ref{Ht}) is
anti-symmetric under time reversion $t\to -t$, i.e., $H(t)=-H(-t)$.
Then the time evolution operators $U(t,0)$ and $U(-t,0)$ obey the
same equation of motion. Moreover, they  obviously are identical and
equal to $\mathbf{1}$ at $t=0$. The equalities $U(\infty,0) =
U(-\infty,0) = U^\dagger(0,-\infty)$ follow immediately, the last
one from unitarity.  Consequently, we find
\begin{equation}
U(\infty,-\infty) = U(\infty,0) U(0,-\infty)
                  = U^\dagger(0,-\infty) U(0,-\infty)
                  = \mathbf{1},
\end{equation}
which implies that at long times, the system will evolve back to its
initial state.

This ideal back-evolution relates our problem to the Loschmidt echo
which has been employed for testing the sensitivity of a ``chaotic''
quantum system on weak perturbations \cite{Jalabert01}.  In the
present case, the small parameter is $\delta\phi = \phi-\pi/2$ which
corresponds to the perturbation Hamiltonian $\Sigma= -2 g \cos(\Om
t)\sin(\delta\phi)\sx$.  Note however, that the present system does
not exhibit any sensitive exponential dependence on the
perturbation.
%

\section{Discussion and summary}

Our study of diabatic level crossing in a system subject to a
time-dependent dipole force revealed two intriguing features. First,
the probability for non-adiabatic transitions is not simply a
monotonous function of the coupling strength but exhibits several
maxima and minima.  In particular, it vanishes for zero coupling and
equals unity if the relation $g/\sqrt{\hbar V} = [2\ln(2)/\pi]^{1/2}
\approx 0.664$ is fulfilled.  This is in contrast to the standard
Landau-Zener problem where the extreme cases require a vanishing or
an infinite interaction strength.
Second, we found that the phase $\phi$ of the dipole field has a
significant influence on the transition probability which is
proportional to $\cos^2\phi$.
The combination of both effects enables one to steer the system
towards the one or the other final state.  In turn, it is also
possible to use the setup as a diagnostic tool for an unknown phase of
a radiation field.
The fact that the results of the transfer matrix approach are only
valid if the duration $\tau_\mathrm{LZ}$ of a single Landau-Zener
transition is sufficiently small, provides a further application.
Measuring the frequency at which the approximation breaks down,
allows one to determine $\tau_\mathrm{LZ}$.
%
%

A straightforward physical realization of our set-up is naturally
provided by spin-$\frac{1}{2}$ systems in time-dependent magnetic
fields. Moreover, one could think of experiments with effective
low-spin systems such as molecular complexes \cite{Sorace03,
Wernsdorfer04}. Indeed, the recent resonant-photon absorption
experiments on the effectively spin-$\frac{1}{2}$ molecular complex
$V_{15}$ could also be performed with a sweep of the magnetic field
that is called $B_{0}$.
A further possible realization is given by Rydberg atoms in the
vicinity of a crossing of the highest Stark level in the $n$th
manifold of the atom and the lowest Stark level of the $n+1$st
manifold \cite{Rubbmark81,Pillet84}.  In this case, the energies are
swept by the {\em dc} Stark effect and the interaction is driven by
a harmonically varying microwave field.  Then, the strong
suppression of non-adiabatic transitions can be tested
experimentally. Particularly promising in this respect would be
variants of microwave ionization experiments of Rydberg atoms that
are based on a mechanism of multiple Landau-Zener transitions to
higher and higher Stark manifolds \cite{Pillet84}.

To summarize, we have studied Landau-Zener transitions in a
two-level atom subject to a harmonically time-dependent, Rabi-like
interaction. As a main difference to the standard Landau-Zener
problem, we find that by tuning the coupling strength within a
relatively small range, it is possible to continuously change the
transition probability from zero to unity.  This behaviour can be
explained within a transfer matrix approach, which provides reliable
results provided that the driving frequency is sufficiently large.
Moreover, this analytical approach allows one to determine the
influence of the phase relation between the diabatic energy crossing
and the dipole field.  It revealed  that the transition probability
is proportional to $\cos^2\phi$ and therefore will vanish for
$\phi=\pi/2$. The latter result was also shown by analyzing the
underlying time-reversal symmetry.  The sensitive phase dependence
can be exploited both for steering the system towards a particular
state and for measuring an a priori unknown phase relation.

\ack

This work has been supported by the Freistaat Bayern via the quantum
information initiative ``Quantum Information Highway A8''. KS and MW
were supported by the Grant-in-Aid for Scientific Research of
Priority Area from the Ministry of Education, Sciences, Sports,
Culture and Technology of Japan (No. 14077216).


\end{document}